\documentclass[amsmath,amssymb,pra,aps,showpacs,twocolumn,10pt,superscriptaddress,tightenlines,floatfix,nofootinbib]{revtex4-2}
\usepackage[colorlinks=true,citecolor=blue,linkcolor=red,urlcolor=blue]{hyperref}
\usepackage[dvipdfmx]{graphicx}
\usepackage{dcolumn}
\usepackage{bm}
\usepackage[utf8]{inputenc}
\usepackage[T1]{fontenc}
\usepackage{lmodern}
\usepackage{cleveref}
\usepackage{color}
\usepackage{amsmath}
\usepackage{amssymb}
\usepackage{amsfonts}
\usepackage{latexsym}
\usepackage{multirow}
\usepackage{booktabs}
\usepackage{comment}
\usepackage{ulem}
\usepackage{bm}

\usepackage{setspace}
\usepackage{here}

\begin{document}
\setlength\textfloatsep{6pt}

\title{Quantum phase transitions in the multiphoton Jaynes-Cummings-Hubbard model}

\author{Hiroo Azuma}
\affiliation{Global Research Center for Quantum Information Science, National Institute of Informatics, 2-1-2 Hitotsubashi, Chiyoda-ku, Tokyo 101-8430, Japan}
\email{zuma@nii.ac.jp}
\author{William J. Munro}
\affiliation{Okinawa Institute of Science and Technology Graduate University, Onna-son, Okinawa 904-0495, Japan}
\author{Kae Nemoto}
\affiliation{Okinawa Institute of Science and Technology Graduate University, Onna-son, Okinawa 904-0495, Japan}
\affiliation{Global Research Center for Quantum Information Science, National Institute of Informatics, 2-1-2 Hitotsubashi, Chiyoda-ku, Tokyo 101-8430, Japan}

\date{\today}

\begin{abstract}
We explore quantum phase transitions in the multiphoton Jaynes-Cummings-Hubbard model (JCHM). Using the mean-field approximation, we
demonstrate that the multiphoton JCHM exhibits
quantum phase transitions
between the Mott insulator (MI) phase, the superfluid phase, and
an additional phase we refer to as
the forbidden phase.
The multiphoton JCHM MI phases are classified according to a conserved quantity associated with the total number of excited atoms and photons.
When this conserved quantity diverges toward positive infinity,
this system
enters
the forbidden phase.
By analyzing the system, we observe MI, superfluid, and forbidden phases in both the single- and two-photon JCHMs,
although the MI phases in the two-photon case are confined to subspaces with small values of the conserved quantity.
In contrast, only
the superfluid and
forbidden
phases
appear
in the three- and four-photon
JCHMs, with no MI phase observed.
\end{abstract}

\keywords{Quantum phase transitions, Jaynes-Cummings-Hubbard model}
\maketitle


\section{\label{section-Introduction}Introduction}
A phase transition is a fundamental phenomenon representing a change between two distinct states of a physical system.
Typically, one can distinguish these states by the behavior of a specific physical quantity that varies in response to an external control parameter, such as temperature or pressure.
The classical example is water, whose phase diagram contains three regions---solid ice, liquid water, and gaseous vapor \cite{Honerkamp2012}.

Despite its familiarity, no exact mathematical model is known that fully reproduces the phase transitions of water within the framework of statistical mechanics.
A tractable example is the two-dimensional Ising model, which offers the simplest theoretical description of ferromagnetism and exhibits a classical phase transition at a critical temperature \cite{Toda1992}.
Classical phase transitions arise due to thermal fluctuations in the system's temperature or pressure.

In contrast, quantum phase transitions occur at zero or near-zero temperature and are driven by changes in non-thermal parameters appearing in the system's Hamiltonian.
These transitions are governed by quantum rather than classical fluctuations.
Notable examples include the magnetic field-induced transitions in quantum Hall systems or the charging energy-driven transitions in Josephson junction arrays \cite{Sondhi1997}.
With rapid progress in
low-temperature
experimental techniques, quantum phase transitions have become an active area of theoretical and experimental investigation.

One important instance is the superfluid-insulator transition \cite{Vojta2003}, particularly the superfluid-Mott insulator transition observed in ultracold atoms confined in optical lattices \cite{Greiner2002}.
These interacting bosonic gases in a lattice potential \cite{Schwabl2006,Paredes2004}
can be described by the Bose-Hubbard model \cite{Fisher1989,Oosten2003,Alet2004}.

The Jaynes-Cummings-Hubbard model (JCHM) offers another platform for studying quantum phase transitions.
It consists of an array of coupled high-Q microcavities, each containing a two-level atom interacting with a single-mode photonic field \cite{Hartmann2006}.
Locally, each cavity is governed by the Jaynes-Cummings model (JCM), while inter-cavity photon tunneling introduces coupling between sites.
Since the theoretical prediction of a Mott insulator (MI) phase of coupled atom-photon excitations in the JCHM \cite{Angelakis2007},
the model has been extensively studied using a range of techniques, including mean-field theory, strong-coupling expansions, Monte Carlo simulations, and density matrix renormalization group (DMRG) methods.

Introducing the superfluid order parameter $\psi=\langle a\rangle$, the expectation value of the photonic annihilation operator,
Refs.~\cite{Greentree2006,Quach2009} used a mean-field approximation to classify quantum phases in the JCHM based on the value of $\psi$ \cite{Lavis2015}.
A non-zero $\psi$ indicates the system is in the superfluid phase, while $\psi=0$ corresponds to the MI phase, enabling the construction of a phase diagram. Beyond mean-field theory, Refs.~\cite{Mering2009,Schmidt2009} studied the JCHM in the strong-coupling limit.
However, exact and general solutions of the JCHM remain an open question.

To investigate critical behavior, large-scale quantum Monte Carlo simulations and DMRG methods have been employed \cite{Hohenadler2011,Rossini2007}.
Reference~\cite{Hohenadler2011} found that the dynamical critical exponent obtained from Monte Carlo simulations differs from that predicted by mean-field theory.
Reference~\cite{Rossini2007} predicted the emergence of a polaritonic glassy phase using the DMRG method.
Additionally, Makin {\it et al}. \cite{Makin2008} examined finite-sized JCHMs (up to six sites) on various lattices, including one-dimensional chains and two-dimensional square and honeycomb lattices under periodic boundary conditions.
They constructed phase diagrams for each topology and compared them with mean-field results that neglect the global lattice structure.
Similarities between the phase diagrams of the JCHM and the Bose-Hubbard model were noted in Ref.~\cite{Tomadin2010}.

As mentioned, the JCHM can be physically realized using arrays of coupled high-Q cavities described by the JCM \cite{Jaynes1963,Shore1993}.
A natural extension is the multiphoton
JCM, with Felicetti {\it et al}. \cite{Felicetti2018} proposing a realization of the two-photon quantum Rabi model (QRM) using superconducting quantum interference devices (SQUIDs) or trapped ions \cite{Felicetti2015,Puebla2017}.
This is significant because the two-photon JCM can be derived from the two-photon QRM via the rotating-wave approximation.

In this work, we investigate quantum phase transitions in multiphoton JCHMs at zero temperature using the mean-field approximation.
We classify the MI phases based on a conserved quantity $L$, defined as the total number of excited atoms and photons \cite{Greentree2006}.

This paper is organized as follows.
Section~\ref{section-JCHM-multiphoton-definition} introduces the Hamiltonian of the multiphoton JCHM
followed by Sec.~\ref{section-implementation-multiphoton-JCHM} which outlines a possible experimental implementation of the two-photon JCHM.
Then, Sec.~\ref{section-mean-field-approximation} presents the superfluid order parameter and the mean-field approximation.
In Sec.~\ref{section-numerical-calculations}, we perform numerical simulations of the multiphoton JCHM under various system parameters and examine their effects on the order parameter.
Section~\ref{section-intuitive-explanations} discusses why the MI phases with $L=1$ and $L=3$ do not emerge in the two-photon JCHM,
and explains the absence of MI phases with finite $L$ in the three- and four-photon cases.
Finally, Sec.~\ref{section-Conclusion} summarizes our conclusions and outlines potential directions for future research.

\section{\label{section-JCHM-multiphoton-definition}The Hamiltonian of the multiphoton JCHM}
The multiphoton JCHM  can be represented by the Hamiltonian \cite{Greentree2006,Felicetti2018,Felicetti2015,Puebla2017}
\begin{equation}
\hat{H}_{\mbox{\scriptsize mpJCH}}
=
\hat{H}_{\mbox{\scriptsize mpJC}}
+
\hat{H}_{\mbox{\scriptsize hop}}
+
\hat{H}_{\mbox{\scriptsize chem}},
\label{multiphoton-JCHM-Hamiltonian-0}
\end{equation}
where
\begin{eqnarray}
\hat{H}_{\mbox{\scriptsize mpJC}}
&=&
\sum_{i=1}^{N}
\hat{H}_{i,\mbox{\scriptsize mpJC}}, \nonumber \\
\hat{H}_{i,\mbox{\scriptsize mpJC}}
&=&
\hbar \Omega
\hat{\sigma}_{i,+}\hat{\sigma}_{i,-}
+
\hbar\omega
\hat{a}_{i}^{\dagger}\hat{a}_{i}
+
\hbar\beta
(
\hat{\sigma}_{i,+}\hat{a}_{i}^{l}+h.c.
), \nonumber \\
\hat{H}_{\mbox{\scriptsize hop}}
&=&
-
\hbar\kappa
\sum_{<i,j>}
\hat{a}_{i}^{\dagger}\hat{a}_{j}, \nonumber \\
\hat{H}_{\mbox{\scriptsize chem}}
&=&
-
\hbar \mu
\sum_{i=1}^{N}
\hat{L}_{i},
\label{multiphoton-JCHM-Hamiltonian-1}
\end{eqnarray}
with $N$ denoting the total number of sites on the lattice, while $\sum_{<i,j>}$ is a summation for nearest neighbors of the lattice.
Here,
$\hat{L}_{i}
=
l\hat{\sigma}_{i,+}\hat{\sigma}_{i,-}+\hat{a}_{i}^{\dagger}\hat{a}_{i}$
for $l=1,2,3,...$.
Further $\hat{a}_{i}$ and $\hat{a}_{i}^{\dagger}$ are the photonic annihilation and creation operators of the $i$-th site, $\hat{\sigma}_{i,+}$ and $\hat{\sigma}_{i,-}$ are the atomic raising and lowering operators of the $i$-th site, while $\kappa$ is the strength of the photon hopping.
For a
one-dimensional (1D)
spin chain,
the Hamiltonian can be simplified,
with the hopping term $\hat{H}_{\mbox{\scriptsize hop}}$ being rewritten as
$
-
\hbar\kappa
\sum_{i=1}^{N-1}
(\hat{a}_{i}^{\dagger}\hat{a}_{i+1}
+
\hat{a}_{i+1}^{\dagger}\hat{a}_{i})
$.
In the single-photon JCHM, the total number of excited atoms and photons is conserved with that quantity being an eigenvalue of the operator $\sum_{i=1}^{N}(\hat{\sigma}_{i,+}\hat{\sigma}_{i,-}+\hat{a}_{i}^{\dagger}\hat{a}_{i})$.
For the $l$-photon JCHM, the operator $\sum_{i=1}^{N}\hat{L}_{i}$ corresponds to this conserved quantity.
Accordingly, to treat the grand canonical ensemble of the model, we introduce the chemical potential $\mu$ in
$\hat{H}_{\mbox{\scriptsize chem}}$ \cite{footnote1}.
The term $\hat{H}_{i,\mbox{\scriptsize mpJC}}$ is the Hamiltonian of the multiphoton JCM at the $i$-th site, where
$\Omega/(2\pi)$ denotes the frequency between atomic excited and ground states, $\omega/(2\pi)$ denotes the frequency of the photons, while $\beta$ denotes the strength of interaction between the atom and $l$ photons.

\section{\label{section-implementation-multiphoton-JCHM}The two-photon JCHM using
Josephson junctions}

\begin{figure}[b]
\begin{center}
\includegraphics[width=0.7\linewidth]{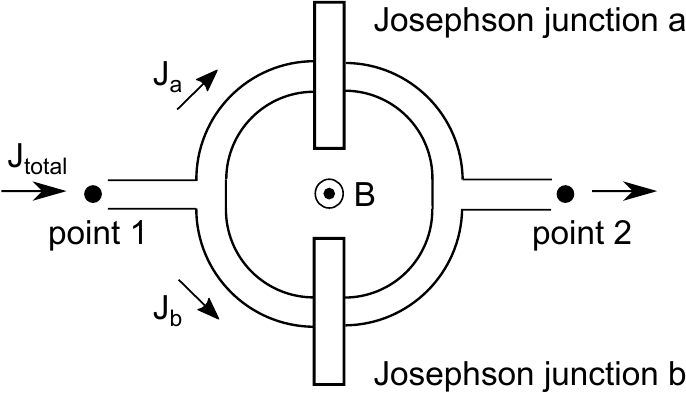}
\end{center}
\caption{
Schematic illustration of a superconducting circuit implementing the two-photon
JCM.
Two Josephson junctions are arranged in parallel within a superconducting loop, with an external magnetic field applied through the interior of the loop.
The left- and right-circulating persistent current states serve as an effective two-level atom, while the magnetic field acts analogously to the cavity field.
}
\label{figure01}
\end{figure}

Let us briefly outline how the two-photon JCHM can be implemented experimentally.
We begin by deriving the two-photon Jaynes-Cummings interaction from a circuit composed of Josephson junctions.
As illustrated in Fig.~\ref{figure01}, we consider a configuration with two Josephson junctions arranged in parallel, and apply an external magnetic field
$\mbox{\boldmath $B$}$
threading the circuit
\cite{Felicetti2018,Azuma2025,Kittel1996}.
We denote the phase differences of wave functions between points $1$ and $2$ along the paths of Josephson junctions $a$ and $b$
by
\begin{equation}
\delta_{a}
=
\delta_{0}
-
\frac{e}{\hbar c}\Phi,
\quad
\delta_{b}
=
\delta_{0}
+
\frac{e}{\hbar c}\Phi,
\end{equation}
respectively, where $\delta_{0}$ is a constant and $\Phi$ is the magnetic flux of $\mbox{\boldmath $B$}$.
We denote the total current and currents along junctions $a$ and $b$ as $J_{\text{total}}$, $J_{a}$, and $J_{b}$, respectively.
The currents $J_{a}$ and $J_{b}$ are induced by the Josephson effect, meaning
\begin{equation}
J_{\text{total}}
=
J_{a}+J_{b} \nonumber \\
=
2J_{0}\sin\delta_{0}\cos\frac{e\Phi}{\hbar c},
\end{equation}
where $J_{0}$ is the maximum zero-voltage current.
Next,
\begin{equation}
J_{0}
\propto
\sigma_{z}
=
|L\rangle\langle L|
-
|R\rangle\langle R|,
\end{equation}
where $|L\rangle$ and $|R\rangle$ are the left- and right-circulating persistent currents.
We regard $\{|L\rangle,|R\rangle\}$ as a two-level system.
By applying a bias to the Josephson junctions, we can transform the basis vectors of the two-level system,
$
|0\rangle
=
(1/\sqrt{2})(|L\rangle +|R\rangle)
$
and
$
|1\rangle
=
(1/\sqrt{2})(|L\rangle -|R\rangle)
$.

Assuming that the magnetic field $\mbox{\boldmath $B$}$ is represented as a coherent state $|\alpha\rangle$ with phase $\alpha$,
we have
$
\Phi
\propto
a+a^{\dagger}
$
and obtain
\begin{equation}
J_{\text{total}}
\propto
\sigma_{x}\cos(a+a^{\dagger}) \nonumber \\
=
\sigma_{x}+\frac{1}{2}\sigma_{x}(a+a^{\dagger})^{2}+... ,
\end{equation}
where the second term
in the expansion
is the one we are really interested in.
Applying the rotating-wave approximation
leads to
the two-photon JCM $\sigma_{+}a^{2}+\sigma_{-}(a^{\dagger})^{2}$.

\begin{figure}[b]
\begin{center}
\includegraphics[width=1.0\linewidth]{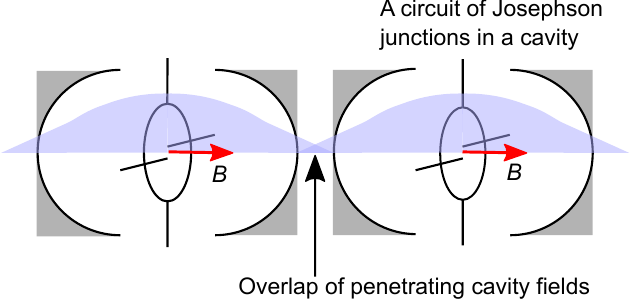}
\end{center}
\caption{
Two adjacent cavities in a one-dimensional lattice.
The wave function of the cavity field extends through the Josephson junction circuit and penetrates the cavity mirrors as tunneling or evanescent light.
}
\label{figure02}
\end{figure}

Next, we construct the two-photon JCHM from the circuits of the Josephson junctions
as shown in Fig.~\ref{figure02} \cite{Hartmann2006}.
It shows two cavities next to each other in the one-dimensional lattice.
The magnetic field $\mbox{\boldmath $B$}$ is induced when the electric current flows along the circuit in each cavity
and it
evolves into a standing wave of the cavity field.
Then, the wave function of the cavity field penetrates the mirror of the cavity as tunneling or evanescent light.
Finally, the adjacent tunneling wave functions or evanescent fields couple in the region of their overlap
and generate the interaction term in the form
$-\hbar\kappa(\hat{a}_{i}^{\dagger}\hat{a}_{j}+\hat{a}_{j}^{\dagger}\hat{a}_{i})$
that is included in the Hamiltonian $\hat{H}_{\mbox{\scriptsize hop}}$ of Eq.~(\ref{multiphoton-JCHM-Hamiltonian-1}),
where $i$ and $j$ denote the neighboring sites of the lattice \cite{Nummemkamp2011}.

Now, we address experimental parameters.
Estimation of a value of $\kappa$, that is, the strength of inter-cavity hopping, is a difficult problem for realistic experiments.
Reference~\cite{Greiner2002} demonstrated experiments with $\tau_{\mbox{\scriptsize tunnel}}=\hbar/\kappa$
where $\tau_{\mbox{\scriptsize tunnel}}$ represented the tunneling time of atoms that were in a three-dimensional optical lattice potential and in a Bose-Einstein condensate.
In those experiments, $\tau_{\mbox{\scriptsize tunnel}}$ was on the order of 2 ms.
Thus, we can expect $\omega_{\kappa}/(2\pi)=\kappa/(2\pi\hbar)\sim$ 80 Hz.
Reference~\cite{Schuster2005} reported an experimental demonstration of a superconducting qubit coupled to a cavity field (the single-photon JCM)
with $\Omega/(2\pi)=$ 6.15 GHz and $\omega/(2\pi)=$ 6 GHz for Eq.~(\ref{multiphoton-JCHM-Hamiltonian-1}).

\section{\label{section-mean-field-approximation}The superfluid order parameter and mean-field approximation}

We now introduce the superfluid order parameter $\psi=\langle \hat{a}_{i}\rangle$, originally formulated within the Ginzburg-Landau theory to describe superconductivity
\cite{Kittel1996,Feynman2006}.
The magnitude of $\psi$ characterizes the amplitude for finding photons at a given site.
For simplicity, we assume $\psi$ to be real.
According to Ref.~\cite{Sheshadri1993},
the most appropriate decoupling approximation of the inter-cavity hopping term for adjacent $i$-th and $j$-th sites is given by
\begin{equation}
\hat{a}_{i}^{\dagger}\hat{a}_{j}
\approx
\langle\hat{a}_{i}^{\dagger}\rangle\hat{a}_{j}
+
\langle\hat{a}_{j}\rangle\hat{a}_{i}^{\dagger}
-
\langle\hat{a}_{i}^{\dagger}\rangle
\langle\hat{a}_{j}\rangle
=
\psi(\hat{a}_{i}^{\dagger}+\hat{a}_{j})
-
\psi^{2}.
\label{mean-fiels-decoupling-0}
\end{equation}
The reason is as follows.
We split the operators $\hat{a}_{i}^{\dagger}$ and $\hat{a}_{j}$ into the mean value and fluctuations as
$\hat{a}_{i}^{\dagger}=\psi+\delta\hat{a}_{i}^{\dagger}$
and
$\hat{a}_{j}=\psi+\delta\hat{a}_{j}$.
Thus, we obtain
\begin{equation}
\hat{a}_{i}^{\dagger}\hat{a}_{j}
\approx
\psi^{2}+\psi(\delta\hat{a}_{i}^{\dagger}+\delta\hat{a}_{j})+O(\delta^{2}).
\end{equation}
This evaluation is consistent with the right-hand side of Eq.~(\ref{mean-fiels-decoupling-0})
because we can rewrite it as
\begin{equation}
(\hat{a}_{i}^{\dagger}+\hat{a}_{j})\psi-\psi^{2}
=
\psi^{2}+\psi(\delta\hat{a}_{i}^{\dagger}+\delta\hat{a}_{j}).
\end{equation}
Hence, we attain the Hamiltonian of the mean-field approximation in the form,
\begin{equation}
\hat{H}_{\mbox{\scriptsize MF}}
=
\hat{H}_{\mbox{\scriptsize mpJC}}
-
z\hbar\kappa\psi(\hat{a}^{\dagger}+\hat{a})
+
z\hbar\kappa\psi^{2}
-
\hbar \mu \hat{L},
\label{Hamiltonian-mean-field-approximation-1}
\end{equation}
where
$z$ denotes the number of nearest neighbors and
\begin{equation}
\hat{H}_{\mbox{\scriptsize mpJC}}
=
\hbar \Omega \hat{\sigma}_{+}\hat{\sigma}_{-}
+
\hbar\omega \hat{a}^{\dagger}\hat{a}
+
\hbar\beta
[\hat{\sigma}_{+}\hat{a}^{l}+\hat{\sigma}_{-}(\hat{a}^{\dagger})^{l}].
\label{Hamiltonian-multiphoton-JC-1}
\end{equation}

Looking at Eqs.~(\ref{Hamiltonian-mean-field-approximation-1}) and (\ref{Hamiltonian-multiphoton-JC-1})
with
$\hat{L}=l\hat{\sigma}_{+}\hat{\sigma}_{-}+\hat{a}^{\dagger}\hat{a}$, we note that the Hamiltonian includes the parameters $\kappa$, $\mu$, $\Omega$, $\omega$, and $\beta$. However, by moving to a scaled time $\beta t$, we can reduce the number of the parameters by one and obtain new parameters $\kappa/\beta$, $\mu/\beta$, $\Omega/\beta$, and $\omega/\beta$. Now, the parameter $\kappa/\beta$ represents the strength of the inter-cavity hopping term. Because the mean-field approximation replaces all effects that a single site receives with an average of the neighboring sites, the hopping term is effective only between the neighboring pair. Thus, the interaction between sites separated by distances of two or more unit lengths is neglected. This means the mean-field approximation depends on the number of the nearest neighbors but not the global topology of connections of the lattice. In other words, our approximation reflects local properties of the lattice but not global ones.
Hence, the above discussion is valid for $\kappa/\beta\ll 1$.
The term $z\hbar\kappa\psi^{2}$ causes the divergence to negative infinity under $|\psi|\to\infty$ if we set $\kappa<0$.
Thus, we only consider cases of $\kappa>0$ to avoid a possible divergence to negative infinity for the minimum value of the total energy as $|\psi|\to\infty$.

Our procedure for determining whether the system undergoes a phase transition is as follows:
we first determine the minimum eigenvalue of $\hat{H}_{\mbox{\scriptsize MF}}$ labelling it $E_{\mbox{\scriptsize min}}$ at that the particular value of $\psi$ denoted by ${\psi}_{E_{\rm min}}$. If ${\psi}_{E_{\rm min}}=0$, we consider that the effect of the inter-cavity hopping term
$-
z\hbar\kappa\psi(\hat{a}^{\dagger}+\hat{a})
+
z\hbar\kappa\psi^{2}$
vanishes and photons are localized at each site.
In this case, the system is identified as being in the MI phase.
By contrast, if ${\psi}_{E_{\rm min}}\neq 0$, we regard the inter-cavity hopping term as effective and consider that photons are transported between neighboring
sites and so are not localized.
This indicates that the system is in the superfluid phase.

We emphasize a cautionary point: it is not necessary to linearize with respect to $\psi$ the terms
$\hbar\omega\hat{a}^{\dagger}\hat{a}$
and
$\hat{\sigma}_{+}\hat{a}^{l}+\hat{\sigma}_{-}(\hat{a}^{\dagger})^{l}$
appearing in the multiphoton Jaynes-Cummings Hamiltonian
$\hat{H}_{\mbox{\scriptsize mpJC}}$ given by Eq.~(\ref{Hamiltonian-multiphoton-JC-1}).
This is because, under the mean-field approximation, we are required to describe
the photonic annihilation and creation operators as $\hat{a}=\psi+\delta\hat{a}$ and $\hat{a}^{\dagger}=\psi+\delta\hat{a}^{\dagger}$ respectively.
Then,
\begin{equation}
\hat{a}^{\dagger}\hat{a}
\to
\psi^{2}
+
\psi(\delta\hat{a}+\delta\hat{a}^{\dagger})
+
\delta\hat{a}^{\dagger}\delta\hat{a}.
\end{equation}
If $|\psi|\gg|\langle\delta\hat{a}\rangle|$
the fluctuation operators can be treated as small, allowing us to neglect the term
$\delta\hat{a}^{\dagger}\delta\hat{a}$.
This leads to a linearized form for 
$\hat{a}^{\dagger}\hat{a}$
as
$\psi(\delta\hat{a}+\delta\hat{a}^{\dagger})$. However,
in the MI phase, the order parameter vanishes
$\psi=0$, $\hat{a}=\delta\hat{a}$, and $\hat{a}^{\dagger}=\delta\hat{a}^{\dagger}$ are satisfied, and we obtain $\hat{a}^{\dagger}\hat{a}=\delta\hat{a}^{\dagger}\delta\hat{a}$. Thus, we cannot rewrite the term $\hat{a}^{\dagger}\hat{a}$
in
a linearized form. 
We can apply the same discussion to $\hat{\sigma}_{+}\hat{a}^{l}+\hat{\sigma}_{-}(\hat{a}^{\dagger})^{l}$, as well.
Hence, we must adopt Eqs.~(\ref{Hamiltonian-mean-field-approximation-1}) and (\ref{Hamiltonian-multiphoton-JC-1}) as the Hamiltonian instead of the linearized one.

\section{\label{section-numerical-calculations}Numerical Simulations}
\begin{figure}[htb]
\begin{center}
\includegraphics[width=1.0\linewidth]{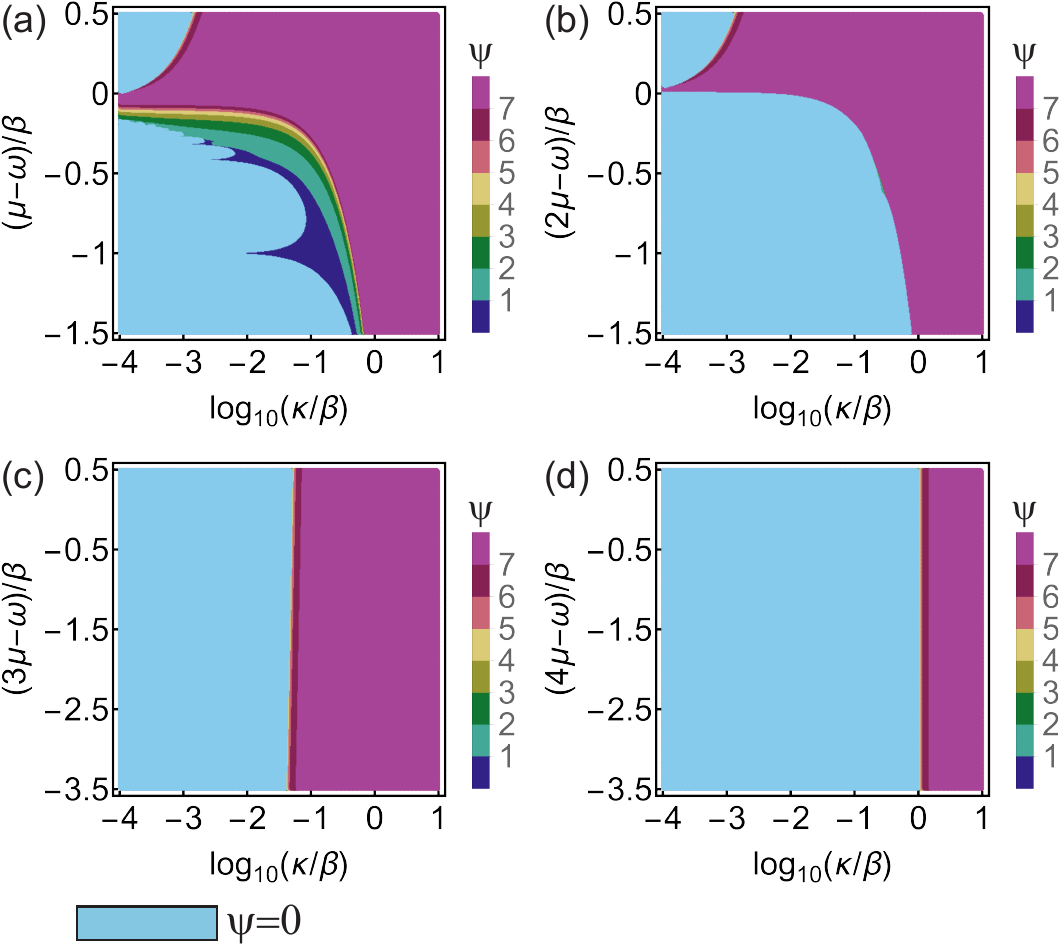}
\end{center}
\caption{Contour plots of $\psi={\psi}_{E_{\rm min}}$ that gives the minimum energy of $\hat{H}_{\mbox{\scriptsize MF}}$ for $l=1$ (a), $2$ (b), $3$ (c), and $4$ (d) respectively with $\Delta=\omega-\Omega=0$, $\mu/\beta=1$, and $z=2$.
The order parameter vanishes ($\psi=0$) in the light blue region.}
\label{figure03}
\end{figure}

\begin{figure}[htb]
\begin{center}
\includegraphics[width=1.0\linewidth]{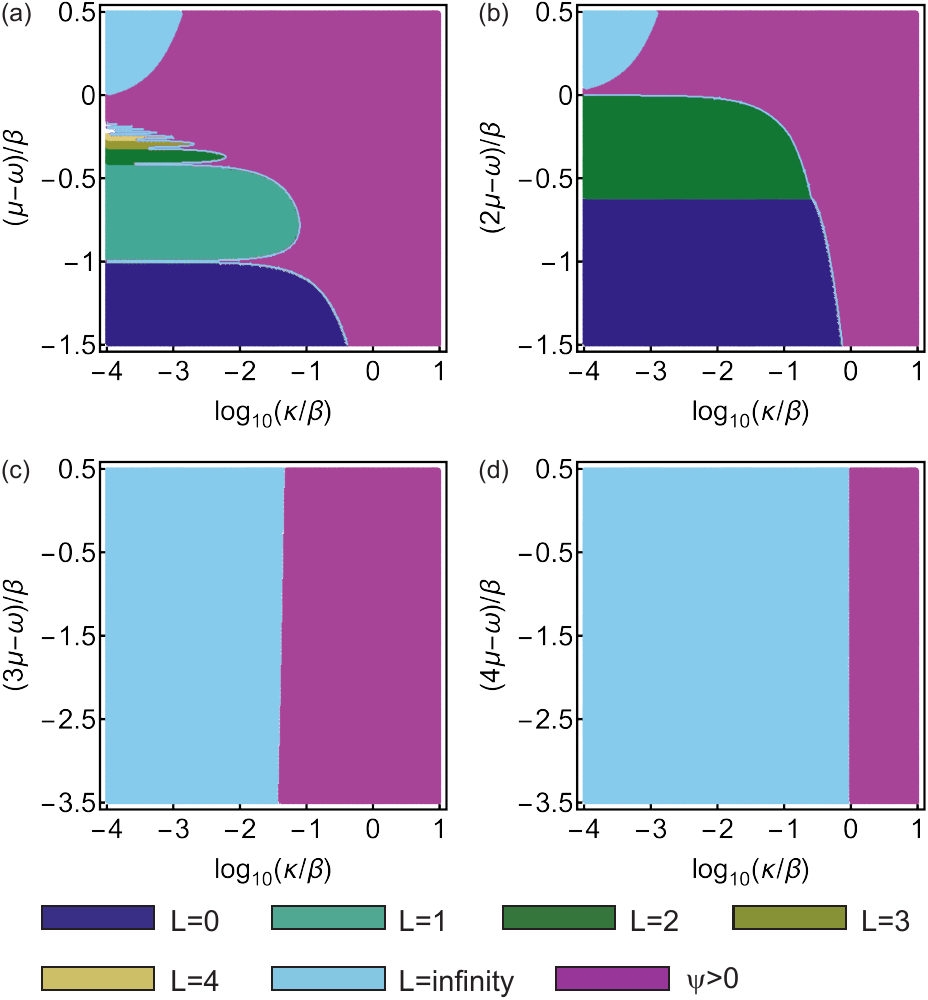}
\end{center}
\caption{Phase diagrams  for $l=1$ (a), $2$ (b), $3$ (c), and $4$ (d) respectively, classified by eigenvalues $L$ with $\Delta=0$, $\mu/\beta=1$, and $z=2$.
The regions of $L=0,1,2,3,4$, and $\infty$ with $\psi=0$ are painted in different colors, respectively.
The white regions represent that $5\leq L<\infty$ with $\psi=0$.
The purple areas represent $\psi>0$.}
\label{figure04}
\end{figure}

In Fig.~\ref{figure03}, we show contour plots of $\psi={\psi}_{E_{\rm min}}$ as a function of $\kappa/\beta$ and $(l\mu-\omega)/\beta$ for various $l$ in the resonance case $\Delta\equiv \omega-\Omega=0$, $\mu/\beta=1$, and $z=2$
corresponding to the 1D spin chain.
Examining Fig.~\ref{figure03}(a), we observe that the region where $\psi=0$ is divided into distinct segments.
For instance, in the case of $l=1$, the lowest light blue region is clearly separated from the upper $\psi=0$
region.
%
In Fig.~\ref{figure04}(a), we paint these parts of $\psi=0$ in different colors for $l=1$.
Figure~\ref{figure03}(b) shows that there are two areas of $\psi=0$ in the graph for $l=2$.
However, in Fig.~\ref{figure04}(b), we observe that the lower area of $\psi=0$ in Fig.~\ref{figure03}(b) is divided into two parts. The values of $\log_{10}(\kappa/\beta)$ at the boundary points of Figs.~\ref{figure03}(c) and (d) do not depend on $(l\mu-\omega)/\beta$ for $l=3$ and $4$, respectively.

Now, let us consider the situation in which the system is in the MI phase. In this
case, where $\psi=0$,
the Hamiltonian simplifies to
\begin{equation}
\hat{H}'_{\mbox{\scriptsize MF}}
=
\hat{H}_{\mbox{\scriptsize mpJC}}
-
\hbar \mu \hat{L}.
\end{equation}
The eigenvalue $L$ is a conserved quantity
as $[\hat{H}_{\mbox{\scriptsize mpJC}},\hat{L}]=0$ for $l$-photon JCHM with $l=1, 2, 3$, and $4$.
This means we can classify the areas of the MI phase according to this conserved quantity.

In Fig.~\ref{figure04}, we classify the regions of the MI phase by the eigenvalue $L$ for $l=1,2,3$, and $4$, respectively.
Examining Fig.~\ref{figure04}(a), we observe that we can divide the region of the MI phase into parts of $L=0,1,2, ..., \infty$ for $l=1$ (the single-photon JCHM).
For $L=0$, the wave function of the lowest energy is given by
$|g\rangle_{\mbox{\scriptsize A}}|0\rangle_{\mbox{\scriptsize P}}$,
where
$\{|g\rangle_{\mbox{\scriptsize A}},|e\rangle_{\mbox{\scriptsize A}}\}$
are atomic ground and excited states and
$\{|n\rangle_{\mbox{\scriptsize P}}:n=0,1,2,...\}$
are the Fock states of the photons.
For $L=1$ and $2$,
the wave functions of the lowest energy are given by
$|\varphi_{-}\rangle=(1/\sqrt{2})(
|g\rangle_{\mbox{\scriptsize A}}|L\rangle_{\mbox{\scriptsize P}}
-
|e\rangle_{\mbox{\scriptsize A}}|L-1\rangle_{\mbox{\scriptsize P}}
)$
but not
$|\varphi_{+}\rangle=(1/\sqrt{2})(
|g\rangle_{\mbox{\scriptsize A}}|L\rangle_{\mbox{\scriptsize P}}
+
|e\rangle_{\mbox{\scriptsize A}}|L-1\rangle_{\mbox{\scriptsize P}}
)$.
This is because an expectation value of the interaction term in  $\hat{H}_{\mbox{\scriptsize mpJC}}$ given by Eq.~(\ref{Hamiltonian-multiphoton-JC-1}) is equal to
$\langle\varphi_{\pm}|\hbar\beta(\hat{\sigma}_{+}\hat{a}+\hat{\sigma}_{-}\hat{a}^{\dagger})|\varphi_{\pm}\rangle =\pm\hbar\beta$. Thus, the energy of $|\varphi_{-}\rangle$ is smaller than that of $|\varphi_{+}\rangle$ if $\beta>0$.
This
is true
for the $l$-photon JCHM with $l=2,3$, and $4$, as well.

In the light blue area of Fig.~\ref{figure04}(a), we obtain $L=M$ and the wave function of the lowest energy is
$(1/\sqrt{2})(
|g\rangle_{\mbox{\scriptsize A}}|M\rangle_{\mbox{\scriptsize P}}
-
|e\rangle_{\mbox{\scriptsize A}}|M-1\rangle_{\mbox{\scriptsize P}}
)$
if we compute $\psi$ with a $(2M+1)$-dimensional Hilbert space spanned by
$\{
|g\rangle_{\mbox{\scriptsize A}}|0\rangle_{\mbox{\scriptsize P}},
|e\rangle_{\mbox{\scriptsize A}}|0\rangle_{\mbox{\scriptsize P}},
|g\rangle_{\mbox{\scriptsize A}}|1\rangle_{\mbox{\scriptsize P}},
|e\rangle_{\mbox{\scriptsize A}}|1\rangle_{\mbox{\scriptsize P}},
...,
\\
|g\rangle_{\mbox{\scriptsize A}}|M-1\rangle_{\mbox{\scriptsize P}},
|e\rangle_{\mbox{\scriptsize A}}|M-1\rangle_{\mbox{\scriptsize P}},
|g\rangle_{\mbox{\scriptsize A}}|M\rangle_{\mbox{\scriptsize P}}
\}$.
As $M\to\infty$, the eigenvalue $L$ diverges to infinity.
Thus, in the light blue region, we have $\psi=0$ and $L=\infty$, indicating that the photons are completely localized at each site, and their number diverges to infinity.
In this situation, the Hamiltonian $\hat{H}_{\mbox{\scriptsize mpJCH}}$ given by Eq.~(\ref{multiphoton-JCHM-Hamiltonian-0}) is nearly equal to
$\hbar(\omega-\mu)L$ as $L\to +\infty$.
Thus, for $\mu-\omega>0$, the total energy of the cavity field diverges to negative infinity and such a state cannot be realized physically.
At first glance, it may appear that the system is energetically attracted to this phase.
However, such a state is never realized in practice, as it would require an infinite number of photons.
Consequently, the system avoids entering this unphysical state.
Due to these considerations, we refer to the light blue region in Fig.~\ref{figure04}(a) as the forbidden phase.
In summary, we regard the system as being in the MI phase when the quantity 
$L$ is finite, and in the forbidden phase when 
$L$ diverges to infinity.

A closer look at Fig.~\ref{figure04}(b) reveals
we can divide the region of $\psi=0$ into three parts, $L=0,2$, and $\infty$ for $l=2$ (the two-photon JCHM).
This indicates the presence of all three phases:
the MI, superfluid, and forbidden phases. For $L=0,2$, and $\infty$, the wave functions of the lowest energy are given by
$|g\rangle_{\mbox{\scriptsize A}}|0\rangle_{\mbox{\scriptsize P}}$,
$c_{0}|g\rangle_{\mbox{\scriptsize A}}|2\rangle_{\mbox{\scriptsize P}}
+c_{1}|e\rangle_{\mbox{\scriptsize A}}|0\rangle_{\mbox{\scriptsize P}}$
and
$(1/\sqrt{2})(|g\rangle_{\mbox{\scriptsize A}}|M\rangle_{\mbox{\scriptsize P}}
-
|e\rangle_{\mbox{\scriptsize A}}|M-2\rangle_{\mbox{\scriptsize P}})$
with $M\to\infty$, respectively,
where the values of $c_{0}$ and $c_{1}$ depend on $\beta/\omega$.
Further in
Fig.~\ref{figure04}(b), the boundary point between the dark blue and dark green regions is given by
$(2\mu-\omega)/\beta=-0.6180$ for $\log_{10}(\kappa/\beta)=-4$.
The smallest value of $(2\mu-\omega)/\beta$ for the light blue area with $\log_{10}(\kappa/\beta)=-4$ is given by $0.00116$.

Now, exploring Figs.~\ref{figure04}(c) and (d), we oberve that $L$ diverges to infinity in the region of $\psi=0$ for $l=3$ and $4$.
Therefore, for $l=3$ and $4$, we interpret these regions as belonging to the forbidden phase.
The wave functions of the forbidden phase are given by
$(1/\sqrt{2})(|g\rangle_{\mbox{\scriptsize A}}|M\rangle_{\mbox{\scriptsize P}}
-
|e\rangle_{\mbox{\scriptsize A}}|M-3\rangle_{\mbox{\scriptsize P}})$
and
$(1/\sqrt{2})(|g\rangle_{\mbox{\scriptsize A}}|M\rangle_{\mbox{\scriptsize P}}
-
|e\rangle_{\mbox{\scriptsize A}}|M-4\rangle_{\mbox{\scriptsize P}})$
with $M\to\infty$ for $l=3$ and $4$, respectively.

Therefore, we conclude that the MI phases of finite $L=1, 3, 4, ...$ do not appear in Fig.~\ref{figure03}(b) for the two-photon JCHM. Moreover, the MI phase does not arise in Figs.~\ref{figure03}(c) and (d) for the three- and four-photon JCHMs. We give intuitive explanations of those observations in Sec.~\ref{section-intuitive-explanations}.

So far, we have used the mean-field approximation to investigate quantum phase transitions in the multiphoton JCHMs. To assess the reliability of this approach, we now compare its results with those obtained from the strong-coupling limit approximation. Reference~\cite{Mering2009} proposed an effective strong-coupling model for analyzing the phase boundaries of the single-photon JCHM. According to the results in Ref.~\cite{Mering2009}, the equations for the phase boundaries corresponding to
$L=0, 1$, and $2$ are given by
\begin{widetext}
\begin{eqnarray}
(\mu_{L}^{+}-\omega)/\beta
&=&
(\sqrt{L}-\sqrt{L+1})
-
\frac{\kappa}{2-\delta_{L,0}}(\sqrt{L}+\sqrt{L+1})^{2}
\quad\quad
\mbox{for $L=0,1,2$,} \nonumber \\
(\mu_{L}^{-}-\omega)/\beta
&=&
(\sqrt{L-1}-\sqrt{L})
+
\frac{\kappa}{2-\delta_{L,1}}(\sqrt{L}+\sqrt{L-1})^{2}
\quad\quad
\mbox{for $L=1,2$,}
\label{upper-lower-bounds-mu}
\end{eqnarray}
\end{widetext}
where $\mu_{L}^{+}$ and $\mu_{L}^{-}$ represent the upper and lower phase boundaries of the eigenvalue $L$, respectively.

\begin{figure}
\begin{center}
\includegraphics[width=1.0\linewidth]{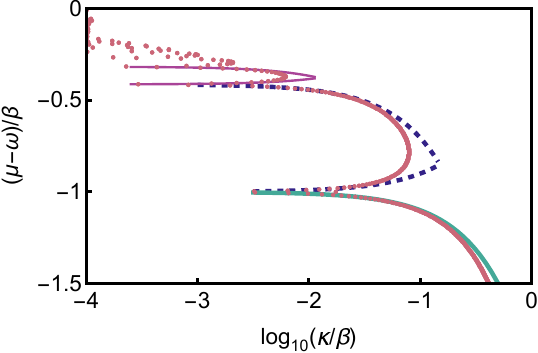}
\end{center}
\caption{
Phase boundaries for the single-photon JCHM, calculated using both mean-field and strong-coupling limit approximations with
$\Delta=0$, $\mu/\beta=1$ and $z=2$.
The pink points represent the phase boundaries extracted from Fig.~\ref{figure04}(a), obtained using the mean-field approximation.
The thick solid green, thick dashed dark blue, and thin solid purple curves correspond to the phase boundaries for $L=0$, $L=1$, and $L=2$, respectively, based on the strong-coupling limit approximation.
}
\label{figure05}
\end{figure}

In Fig.~\ref{figure05},
we show the phase boundaries of the single-photon JCHM calculated using both the mean-field and strong-coupling limit approximations.
The pink points are extracted from Fig.~\ref{figure04}(a) and represent the results of the mean-field approximation.
The thick solid green, thick dashed dark blue, and thin solid purple curves represent the upper and lower phase boundaries for $L=0$, $L=1$, and $L=2$, respectively.
As seen in Fig.~\ref{figure05}, the results from both approximations agree well when
$\kappa/\beta$ is sufficiently small.
However, for larger $\kappa/\beta$, the thick dashed dark blue and thin solid purple curves exhibit sharp features that are not captured by the mean-field approximation.
Therefore, we conclude that the mean-field approximation is reliable only in the regime of small $\kappa/\beta$.

\begin{figure}[b]
\begin{center}
\includegraphics[width=1.0\linewidth]{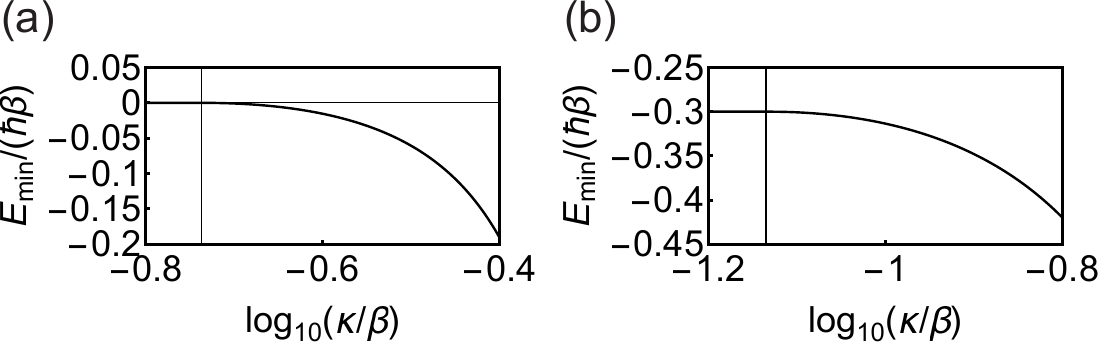}
\end{center}
\caption{Plots of the minimum energies $E_{\mbox{\scriptsize min}}/(\hbar\beta)$ as a function of $\log_{10}(\kappa/\beta)$ for $\Delta=0$, $\mu/\beta=1$, and $z=2$ for the single-photon JCHM ($l=1$).
(a) correspoinds to $(\mu-\omega)/\beta=-1.2$, $L=0$,
while (b) corresponds to $(\mu-\omega)/\beta=-0.7$, $L=1$.
The boundary points between $\psi=0$ and $\psi>0$ are given by $\log_{10}(\kappa/\beta)=-0.737$ and $-1.14$, respectively.}
\label{figure06}
\end{figure}

From the above considerations, we find three phases of the multiphoton JCHMs, the MI, superfluid, and forbidden phases.
Here, we investigate the sharpness of the transitions between these phases.
Figure~\ref{figure06} show plots of the minimum energies as functions of $\log_{10}(\kappa/\beta)$ for $\Delta=0$, $\mu/\beta=1$, $z=2$, and $l=1$ (the single-photon JCHM) with choosing specific values of $(\mu-\omega)/\beta$.
For Figs.~\ref{figure06}(a) and (b), values of $\Delta E_{\mbox{\scriptsize min}}/[(\hbar\beta)\Delta\log_{10}(\kappa/\beta)]$, that is, ratios of differences of $E_{\mbox{\scriptsize min}}/(\hbar\beta)$ to differences of $\log_{10}(\kappa/\beta)$, near the boundary points are
of order $10^{-1}$. Thus, we can regard the transitions as sharp. The values of $|E_{\mbox{\scriptsize min}}|/(\hbar\beta)$ are of order $10^{-1}$ for the phase transitions between the MI and superfluid phases.

We examine how the minimum energy changes near the phase boundaries, as shown in Fig.~\ref{figure06}.
However, as indicated in Fig.~\ref{figure05}, the mean-field approximation becomes less reliable when $\kappa/\beta$ is large.
Therefore, the analyses based on Fig.~\ref{figure06} should be interpreted with caution, as they may not be entirely accurate.
The results derived from Fig.~\ref{figure06} should be regarded as qualitative suggestions rather than definitive conclusions.

\section{\label{section-intuitive-explanations}The presence or absence of MI phases in the multiphoton JCHMs}

In this section, we first provide intuitive explanations for why the MI phases with $L=1$ and $L=3$ do not appear in the two-photon JCHM, as shown in Fig.~\ref{figure04}(b).
Next, we discuss the reasons why the MI phases fail to emerge in the three- and four-photon JCHMs.
To focus on the behavior of the system within the MI and forbidden phases, we assume $\psi=0$.
For the $l$-photon JCHM with $\psi=0$, the dimensionless Hamiltonian
derived from Eqs.~(\ref{Hamiltonian-mean-field-approximation-1}) and (\ref{Hamiltonian-multiphoton-JC-1}) can be expressed as:
\begin{equation}
\tilde{H}^{(l)}
=
\tilde{\Omega}\hat{\sigma}_{+}\hat{\sigma}_{-}
+
\tilde{\omega}\hat{a}^{\dagger}\hat{a}
+
\hat{\sigma}_{+}\hat{a}^{l}
+
\hat{\sigma}_{-}(\hat{a}^{\dagger})^{l}
-
\tilde{\mu}\hat{L},
\end{equation}
where $\tilde{\Omega}=\Omega/\beta$, $\tilde{\omega}=\omega/\beta$, and $\tilde{\mu}=\mu/\beta$. As mentioned in Sec.~\ref{section-mean-field-approximation},
this simplified Hamiltonian is valid for $\kappa/\beta\ll 1$.

Now, let us focus on the case of $l=2$, the two-photon JCHM.
First, consider the state with
$L=0$. It is equal to $|g\rangle_{\mbox{\scriptsize A}}|0\rangle_{\mbox{\scriptsize P}}$ and its dimensionless energy is given by $\tilde{E}^{(2)}_{0}=0$. Second, we consider a state with $L=1$ which can be represented by $|g\rangle_{\mbox{\scriptsize A}}|1\rangle_{\mbox{\scriptsize P}}$ with its dimensionless energy given by $\tilde{E}^{(2)}_{1}=\tilde{\omega}$. This shows that $\tilde{E}^{(2)}_{0}<\tilde{E}^{(2)}_{1}$ holds $\forall\tilde{\omega}>0$ and the phase of $\psi=0$ and $L=1$ cannot be realized.
Thus, the MI phase of $L=1$ does not appear in Fig.~\ref{figure04}(b). Third if we consider a state with $L(\geq l)$ for the $l$-photon JCHM, it can be represented by a superposition of $|g\rangle_{\mbox{\scriptsize A}}|L\rangle_{\mbox{\scriptsize P}}$ and $|e\rangle_{\mbox{\scriptsize A}}|L-l\rangle_{\mbox{\scriptsize P}}$. Taking $\{|g\rangle_{\mbox{\scriptsize A}}|L\rangle_{\mbox{\scriptsize P}},\\
|e\rangle_{\mbox{\scriptsize A}}|L-l\rangle_{\mbox{\scriptsize P}}\}$ as an orthonormal basis,
we describe $\tilde{H}^{(l)}$ in the following $2\times 2$ matrix form:
\begin{equation}
\tilde{H}^{(l)}
=
\left(
\begin{array}{cc}
\tilde{\Omega}+(L-l)\tilde{\omega}-L\tilde{\mu} & \sqrt{L!/(L-l)!} \\
\sqrt{L!/(L-l)!} & L\tilde{\omega}-L\tilde{\mu} \\
\end{array}
\right),
\label{2x2-Hamiltonian-MI-phase-0}
\end{equation}
The eigenvalues of $\tilde{H}^{(l)}$ with $\Delta=\omega-\Omega=0$ and $\tilde{\mu}=1$ are
\begin{equation}
\tilde{E}^{(l)}_{L,\pm}
=
\frac{1}{2}
[-2L+(2L-l+1)\tilde{\omega}
\pm
\sqrt{4L!/(L-l)!+(l-1)^{2}\tilde{\omega}^{2}}].
\label{Eigenvalue-l-L-pm}
\end{equation}
Now by letting $l=2$ and $L=2$ we have
\begin{equation}
\tilde{E}^{(2)}_{2,-}
=
\frac{1}{2}(-4+3\tilde{\omega}-\sqrt{8+\tilde{\omega}^{2}}),
\end{equation}
which is less than zero for $\tilde{\omega}<(1/2)(3+\sqrt{5})\simeq 2.6180$. Thus, if $(2\mu-\omega)/\beta<-0.6180$, the $L=0$ MI phase arises.
By contrast, if $(2\mu-\omega)/\beta>-0.6180$, the $L=2$ MI phase arises. This can be observed Fig.~\ref{figure04}(b). Substituting $l=2$ and $L=3$ into Eq.~(\ref{Eigenvalue-l-L-pm}),
we obtain
\begin{equation}
\tilde{E}^{(2)}_{3,\pm}
=
\frac{1}{2}
(-6+5\tilde{\omega}
\pm\sqrt{24+\tilde{\omega}^{2}}).
\end{equation}
Now $\tilde{E}^{(2)}_{3,-}=\tilde{E}^{(2)}_{2,-}$ holds for $\tilde{\omega}\simeq 1.9215$. Hence, if $(2\mu-\omega)/\beta>0.0785$, the MI phase of $L=3$ could emerge. However, this situation is forbidden
as we will see next.
Let us consider
the large $L$ case.
As
\begin{equation}
\lim_{L\to\infty}\frac{1}{L}\tilde{E}^{(2)}_{L,-}
=
\tilde{\omega}-2,
\end{equation}
$\tilde{E}_{L,-}^{(2)}\to -\infty$ and $\tilde{E}_{L,-}^{(2)}<\tilde{E}_{3,-}^{(2)}$ hold for $\tilde{\omega}<2$ under $L\to\infty$. The forbidden phase must appear for $\tilde{\omega}<2$, that is $(2\mu-\omega)/\beta>0$.
This means the MI phase of $L=3$ cannot arise. On the contrary, the forbidden phase appears.

Finally, there is a similar explanation for why the MI phases cannot be observed in the three- and four-photon JCHMs. Let us set $l=3$ and consider a state whose conserved quantity is equal to $L$.
In such a case we have
\begin{equation}
\lim_{L\to\infty}\frac{\tilde{E}^{(3)}_{L,-}}{L^{3/2}}
=
-1
\quad
\forall\tilde{\omega}.
\end{equation}
Thus, the superposition of $|g\rangle_{\mbox{\scriptsize A}}|L\rangle_{\mbox{\scriptsize P}}$ and $|e\rangle_{\mbox{\scriptsize A}}|L-3\rangle_{\mbox{\scriptsize P}}$ is the lowest energy state with $\tilde{E}_{L,-}^{(3)}\to -\infty$ $\forall\omega$ as $L\to\infty$.
Therefore, we conclude that the forbidden phase appears and the MI phases do not arise in the three-photon JCHM.
A similar behavior is observed in the four-photon JCHM case.

\section{\label{section-Conclusion}Conclusion}

In this work, we investigated the quantum phase transitions of the $l$-photon JCHMs for $l=1,2,3,4$ using the mean-field approximation.
This approach allowed us to construct phase diagrams that clearly reveal the MI, superfluid, and forbidden phases across the multiphoton JCHMs.
We found that all three phases---MI, superfluid, and forbidden---appear
in the single- and two-photon JCHMs phase diagrams.
In contrast, the two-photon JCHM supports MI phases only at $L=0$ and $L=2$.
Remarkably, the three- and four-photon JCHMs exhibit only superfluid and forbidden phases, with the MI phase entirely absent.
The presence of a forbidden phase indicates that the system avoids occupying certain parameter regimes,
even when the chemical potential $\mu$, cavity frequency $\omega$, hopping strength $\kappa$, and coupling $\beta$ are tuned to specific values.
This nontrivial behavior could motivate experimental investigations in parameter regimes where multiphoton interactions dominate and phase transitions exhibit unconventional structure.

\section*{Acknowledgment}
This work was supported by MEXT Quantum Leap Flagship Program Grant No. JPMXS0120351339.


\begin{thebibliography}{99}
\bibitem{footnote1}
Strictly speaking, we must consider $2N$ chemical potentials $\mu_{i,\mbox{\scriptsize A}}$ and $\mu_{i,\mbox{\scriptsize P}}$ for the atoms and the photons at the $i$-th site, respectively, and add a term
$\hbar \sum_{i=1}^{N}
(\mu_{i,\mbox{\scriptsize A}}\hat{\sigma}_{i,+}\hat{\sigma}_{i,-}
+
\mu_{i,\mbox{\scriptsize P}}\hat{a}_{i}^{\dagger}\hat{a}_{i})$
to the Hamiltonian $\hat{H}_{\mbox{\scriptsize mpJCH}}$ instead of $\hbar\mu\sum_{i=1}^{N}\hat{L}_{i}$. However, we assume the common chemical potential $\mu$ for the model for the sake of simplicity.
\end{thebibliography}
\end{document}